\documentclass[conference]{IEEEtran}
\IEEEoverridecommandlockouts
\usepackage{cite}
\usepackage{amsmath,amssymb,amsfonts}
\usepackage{algorithmic}
\usepackage{graphicx}
\usepackage{textcomp}
\usepackage{xcolor}
\usepackage{subcaption}
\usepackage{makecell}
\usepackage{todonotes}
\usepackage{multirow}
\usepackage{verbatim}
\usepackage{todonotes}

\def\BibTeX{{\rm B\kern-.05em{\sc i\kern-.025em b}\kern-.08em
    T\kern-.1667em\lower.7ex\hbox{E}\kern-.125emX}}
    
\begin{document}

%\title{Memristive Memory Array Training  based on Learning Automaton}

%For all other papers 
\IEEEpubid{\makebox[\columnwidth]{ 979-8-3315-0498-4/24/\$31.00 $ \copyright$2024 IEEE \hfill{ \hspace{\columnsep}\makebox[\columnwidth] { }}}} 

\title{In-Memory Learning Automata Architecture using Y-Flash Cell}

%\title{Embedding the  Learning Automata in Memristive Memory Array Cells}
    
%\todo{How about In-Memory Learning Automata Architecture using Y-flash Cell}

\author{
Omar Ghazal$^{\dagger,\star}$,Tian Lan$^\dagger$,Shalman Ojukwu$^\dagger$, Komal Krishnamurthy$^\dagger$, Alex Yakovlev$^\dagger$, Rishad Shafik$^\dagger$\\
$^\dagger$Microsystems Research Group, Newcastle University, UK, 
$^\star$University of Mosul, IRAQ\\
\textrm{\{o.g.g.awf2, t.lan3, s.ojukwu, k.krishnamurthy3, alex.yakovlev, rishad.shafik\}@newcastle.ac.uk}, omargg@uomosul.edu.iq
\\[-3ex]
}%\vspace*{-1mm}

%\and
%\and
%\and
%\and
%\and

\maketitle

\begin{abstract}
The modern implementation of machine learning architectures faces significant challenges due to frequent data transfer between memory and processing units. In-memory computing, primarily through memristor-based analog computing, offers a promising solution to overcome this von Neumann bottleneck. In this technology, data processing and storage are located inside the memory. Here, we introduce a novel approach that utilizes floating-gate Y-Flash memristive devices manufactured with a standard 180 nm CMOS process. These devices offer attractive features, including analog tunability and moderate device-to-device variation; such characteristics are essential for reliable decision-making in ML  applications. This paper uses a new machine learning algorithm, the Tsetlin Machine (TM), for in-memory processing architecture. The TM's learning element, Automaton, is mapped into a single Y-Flash cell, where the Automaton's range is transferred into the Y-Flash's conductance scope. Through comprehensive simulations, the proposed hardware implementation of the learning automata, particularly for Tsetlin machines, has demonstrated enhanced scalability and on-edge learning capabilities.
\end{abstract}


\section{Introduction}
In modern computing systems where the data processing scalability grows alongside the expansion of big data, the Von Neumann bottleneck has become a challenge ~\cite{Abu2024, FeRAM,memristive2023}. The traditional architecture relies on moving data between the memory and processing unit, leading to significant data throughput and energy costs. A new approach called in-memory computing (IMC) addresses these challenges, where data is processed directly within a memory array, eliminating the need for constant data movement between memory and processing units. The IMC has significant advantages, primarily due to the inherent characteristics of the memory cells used in IMC operations, such as having low access time, high cycling endurance, and non-volatility.

\begin{figure}[t]
\centering
\setlength\belowcaptionskip{-1.5\baselineskip}
\includegraphics[width=0.488\textwidth]{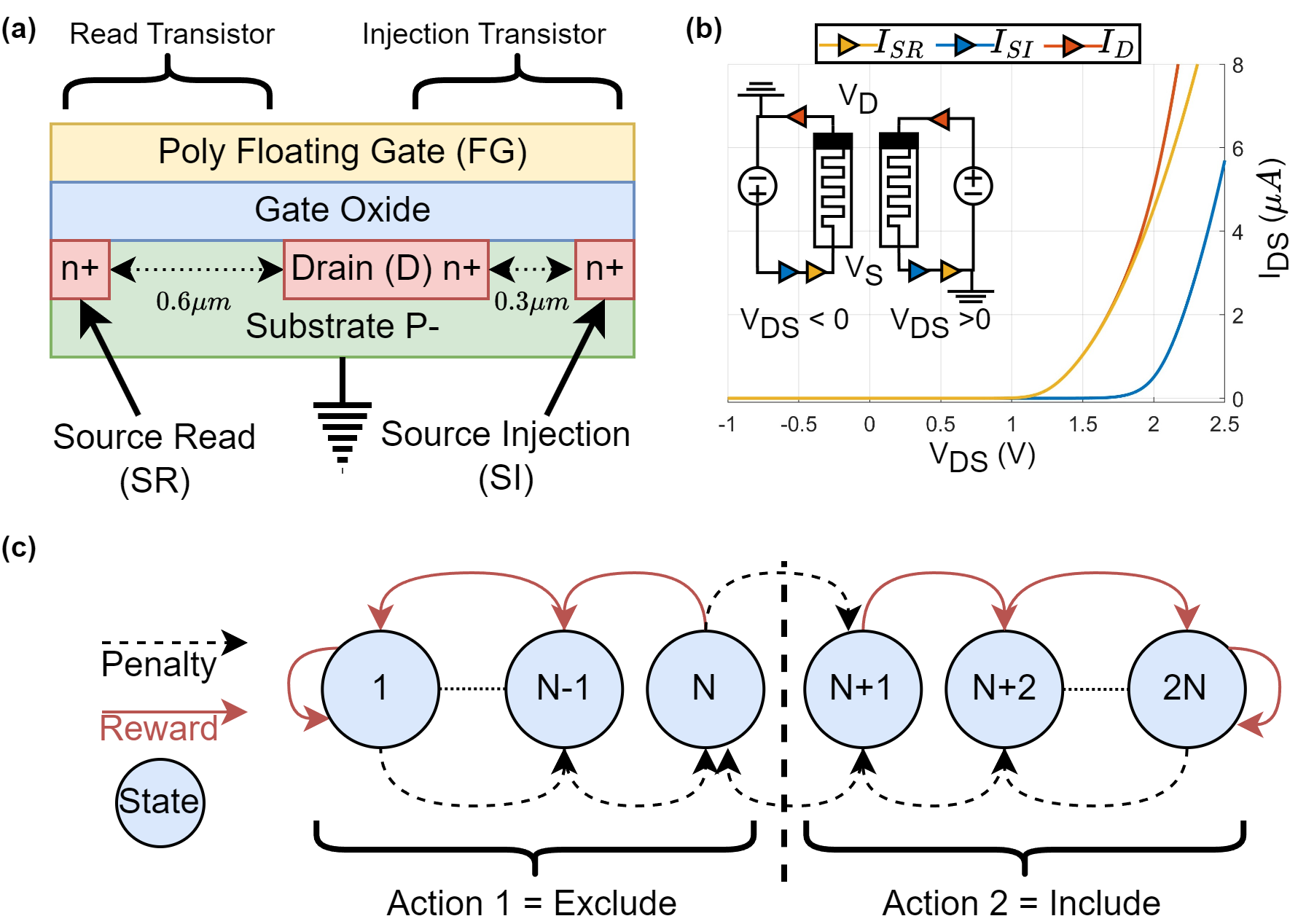}
\caption{Learning Tsetlin Automaton based on a Y-Flash device. (a) Visual representation of the Y-Flash device. (b) Connection of the two sources to form a dual-terminal setup. Demonstrating the device's pronounced self-selection capability through a low positive voltage application. The device state is determined by the current under positive bias (VDS = 2V). (c) Tsetlin Automaton FSM.}
\label{fig1}
\end{figure}

The emerging nonvolatile memory (NVM) devices, like resistive random-access memory (ReRAM) \cite{memristive2023}, phase-change memory (PCM) \cite{Abu2024}, and ferroelectric random-access memory (FeRAM) \cite{FeRAM}, have attracted attention for use as core devices in IMC systems. Aiming to leverage the integration of nonvolatile memory devices with the IMC can significantly improve the performance of computing systems such as Machine learning (ML) algorithms, where processing complex and large volumes of information immediately and efficiently is critical~\cite{Abu2024, FeRAM,memristive2023}. Recent research has explored integrating the ReRAM array with a state-of-the-art ML algorithm, Tsetlin machines (TM) ~\cite{imbue,AsynhImbue}. TM employs a learning Automaton called the Tsetlin Automaton (TA) as its primary learning component, as illustrated in Fig.~\ref{fig1}(c) \cite{2018-granmo-tsetlin}. The TA is responsible for constructing logical propositions that link input-output pairs in classification tasks. The TA behaves similarly to traditional Finite State Machines (FSMs) and incorporates learning mechanisms based on reinforcement learning principles, allowing it to undertake two actions, inclusion or exclusion, based on reward and penalty feedback during training. Upon completion of training, the TA makes a final decision to either include ('1') or exclude ('0') a specific feature. TM forms multiple clauses, each containing a set of TAs generating diverse logical propositions from exact boolean literals (features and their negations). Organized into an architecture of classes, these clauses facilitate the classification algorithm \cite{2018-granmo-tsetlin,redress}.

The structure of TM, based on basic propositional logic, enabled its integration into IMC technology, where the final boolean actions of TAs are mapped onto the ReRAM array \cite{imbue,AsynhImbue}. However, the challenge lies in mapping the multi-state dynamics of TAs onto memristors' analog tunable resistance levels to establish an efficient, low-power IMC system \cite{FSM}. Such systems enable analog computation directly within memory devices. Nevertheless, the actual behavior of memristive devices often does not fully meet the desired specifications. High production yields, numerous conductance states, uniformity, long retention time, and endurance are necessary for developing IMC suitable for machine learning applications \cite{redoxReRAM_Req}. Yet, existing technology requires further refinement to meet these criteria, and it often requires unique materials or processes when integrated into industrial standard complementary metal-oxide-semiconductor (CMOS) processes \cite{meetReqyield,meetreqCMOS}.

Conversely, Y-Flash, NVM floating gate (FG) devices have been developed using CMOS process (see Fig.~\ref{fig1}(a)) \cite{YFLASH2022,YFLASH2019}. The Y-Flash device features a floating gate isolated from its other terminals, allowing it to be charged or discharged. This, in turn, adjusts the threshold of the corresponding transistor to the desired level. The control gate is merged with the drain, significantly reducing the device footprint while allowing for program, erase, and readout operations in a two-terminal configuration as shown in Fig.\ref{fig1}(b) \cite{YFLASH2022,YFLASH2019}. When arranged in a crossbar array, the Y-Flash devices demonstrated a low sneak-path current, eliminating the need for the selector device. Fig.~\ref{fig1}(b) shows the Y-Flash memristor operates in an analog fashion and can be operated in the subthreshold range, resulting in many tunable conductance states and low readout currents $(1nA - 5\mu A)$. Compared to other NVMs, the Y-Flash memristor offers several advantages, including full CMOS process compatibility, low cycle-to-cycle variations, high yield, low power consumption, analog conductance tunability, self-selection, and high retention time \cite{YFLASH2022, YFLASH2019}.

\begin{figure}[t]
\centering
\setlength\belowcaptionskip{-1.5\baselineskip}
\includegraphics[width=0.488\textwidth]{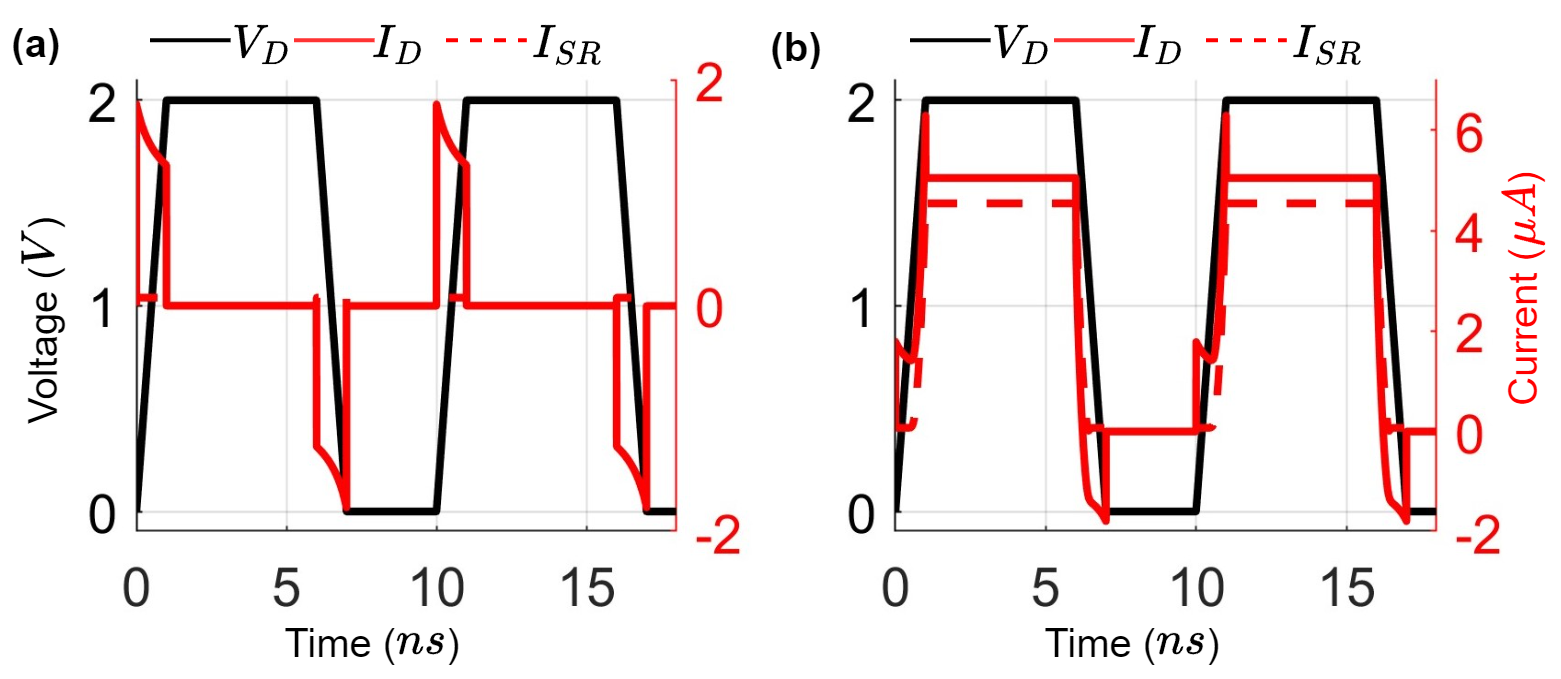}
\caption{The DC IV characteristics of a Y-FLASH device during a reading pulse ($V_D$ = 2V) exhibit distinct behaviors based on its conductance state: (a) In the low conductance state, $I_D$ is $\approx$1nA, and (b) in the high conductance state, $I_D$ increases significantly to $\approx$ 5$\mu$A.}
\label{fig2}
\end{figure}

In addition to the features mentioned previously, scalability is also required for TM, which has an extensive number of TAs that need a memory capable of storing a high data density. These factors make Y-Flash a promising solution for overcoming the limitations of IMC in the context of TM applications.

The main focus of this paper is to explore the practicality and advantages of incorporating the learning Automata, notably TA, into Y-Flash memristive devices. We demonstrate how one Y-Flash cell can be utilized to create up to 40 distinct states of TA, with the ability to reach up to 1000 states through careful tuning. This exploration includes conducting thorough experimental validations via device-to-device (D2D) and cycle-to-cycle (C2C) analyses. We aim to showcase this integration's significant potential to enable a more efficient and adaptable computing approach. The paper contributions are:

\begin{figure}[tbp]
\centering
\setlength\belowcaptionskip{-1.5\baselineskip}
\includegraphics[width=0.488\textwidth]{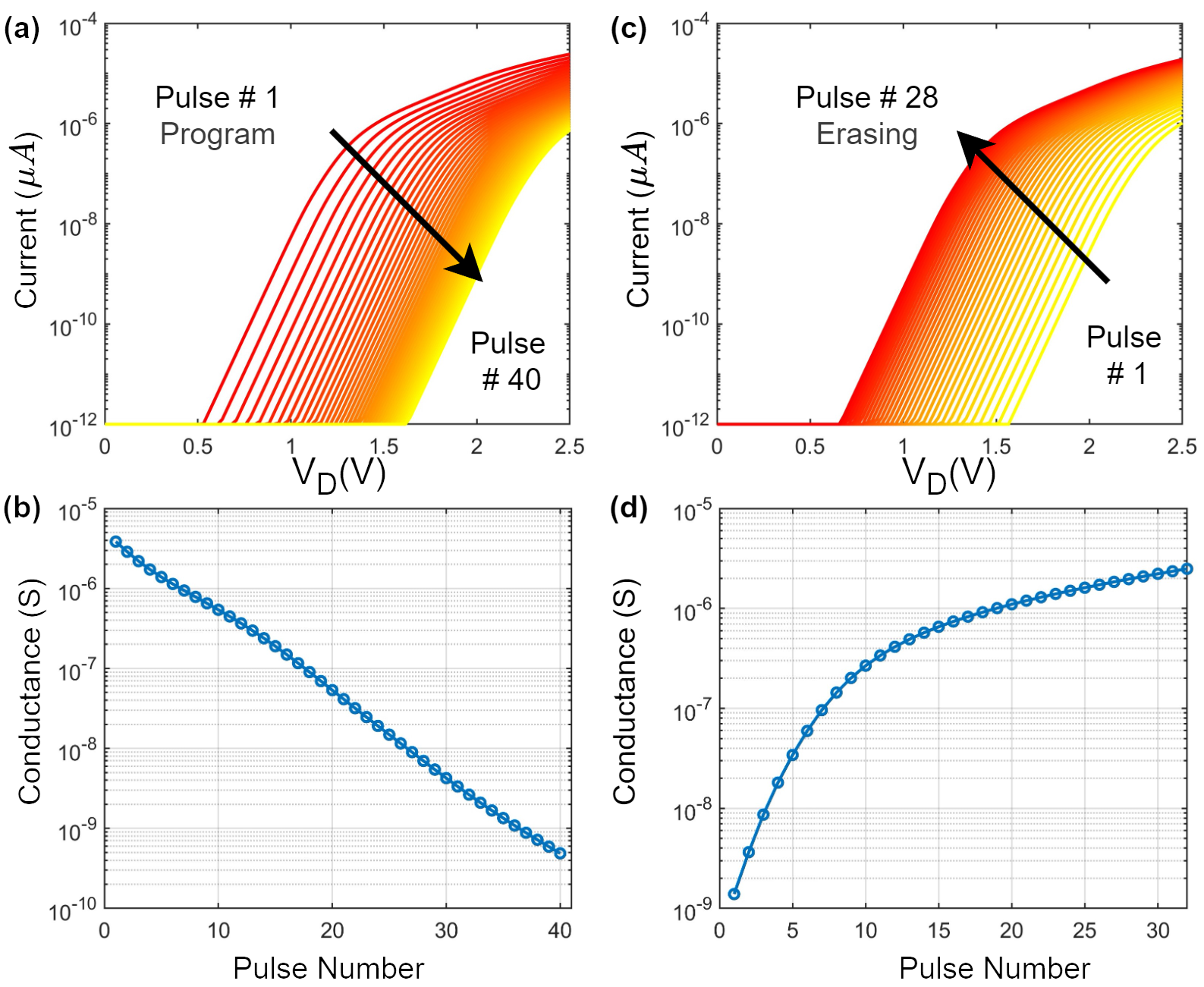}
\caption{ Multi-conductance states achieved through repeated pulses. (a) and (c) $I_D$ vs. $V_D$ for programming and erasing, respectively. (b) and (d) Conductance (at $V_R = 2V$) change with pulse number for programming and erasing, respectively.}
\label{fig3}
\end{figure}

\begin{itemize}
    \item Explore the integration of TA  states with Y-Flash memristive devices for  In-Memory Computing.
    \item Demonstrate the Y-Flash cell's ability to achieve up to 40 distinct states, extendable to 1000 through tuning.
    \item Validate Y-Flash devices' performance, reliability, and scalability for computing applications.
\end{itemize}

\section{Y-Flash Based Tsetlin Automaton Implementation } This section explores the application of encoding the states of TA as conductances utilizing a Y-Flash memristive device. It thoroughly examines the device's characteristics and explains the applied procedural approach for mapping the states of TA into the Y-Flash cell. For this purpose, a compact Y-Flash model is employed, considering both D2D and C2C variations in design \cite{YFLASHMODEL}.

\subsection{Y-Flash Cell characteristics}
Fig.~\ref{fig1}(a) illustrates the Y-Flash device, which consists of two transistors connected in parallel: the read transistor (SR) and the injection transistor (SI). They share a standard drain (D) and a polysilicon floating gate. This device is fabricated using a commercial 180 nm CMOS process \cite{YFLASH2022}. The SR is optimized for low-read voltage operation below 1V, with a longer channel length $0.6\mu m$ and a lower threshold voltage \((V_{th})\)$\approx0.3V$. On the other hand, the SI is designed for high-voltage program/erase operations, with a shorter channel length $0.3\mu m$ and a higher $V_{th}\approx1.5V$. The Y-Flash device can function as a two-terminal memristor by shortening the two sources, SR and SI, see Fig.~\ref{fig1}(b). This allows for simpler addressing, while the three-terminal configuration permits more flexibility in different operation modes, Table~\ref{Y-FLASH's operation modes}. The behavior of the device's current-voltage $(I_{DS} - V_{DS})$ is shown in Fig.~\ref{fig1}(b). Under positive bias conditions $(V_D > 0 \ and\ V_S = 0)$, the $(I_{DS} - V_{DS})$ relation exhibits nonlinearity due to drain-source and floating-gate voltage changes. Conversely, when voltage bias is reversed $(V_D = 0 \ and\ V_S > 0)$, the current flow is negligible as the floating gate is coupled to the drain, effectively switching off the transistors. This unique I-V behavior, in addition to the low reverse current, allows the device to self-select and minimize sneak-path currents in crossbar arrays without requiring a selector device.

\begin{table}[b]
\vspace{-5pt}
\caption{Y-Flash's operation modes.}
\begin{center}
\begin{tabular}{|c|c|c|c|c|}
\hline
\textbf{Operation}&\multicolumn{4}{|c|}{\textbf{Terminals configurations}} \\
\cline{2-5} 
\textbf{Mode} & \textbf{\textit{Drain}}& \textbf{\textit{Source Read}}& \textbf{\textit{Source Injection}}&\textbf{\textit{Substrate}} \\
\hline
Read& $2V$ & $0$ & $0 / Z$ & $0$ \\\hline
Program& $> 4V$ & $0 / Z$ & $0$  & $0$ \\\hline
Erase& $0 / Z$ & $0 / Z$ & $8V$ & $0$ \\\hline
\multicolumn{4}{l}{$Z$: is high impedance or floating.}
\end{tabular}
\label{Y-FLASH's operation modes}
\end{center}
\end{table}

\begin{figure}[tbp]
\centering
\setlength\belowcaptionskip{-1.5\baselineskip}
\includegraphics[width=0.5\textwidth]{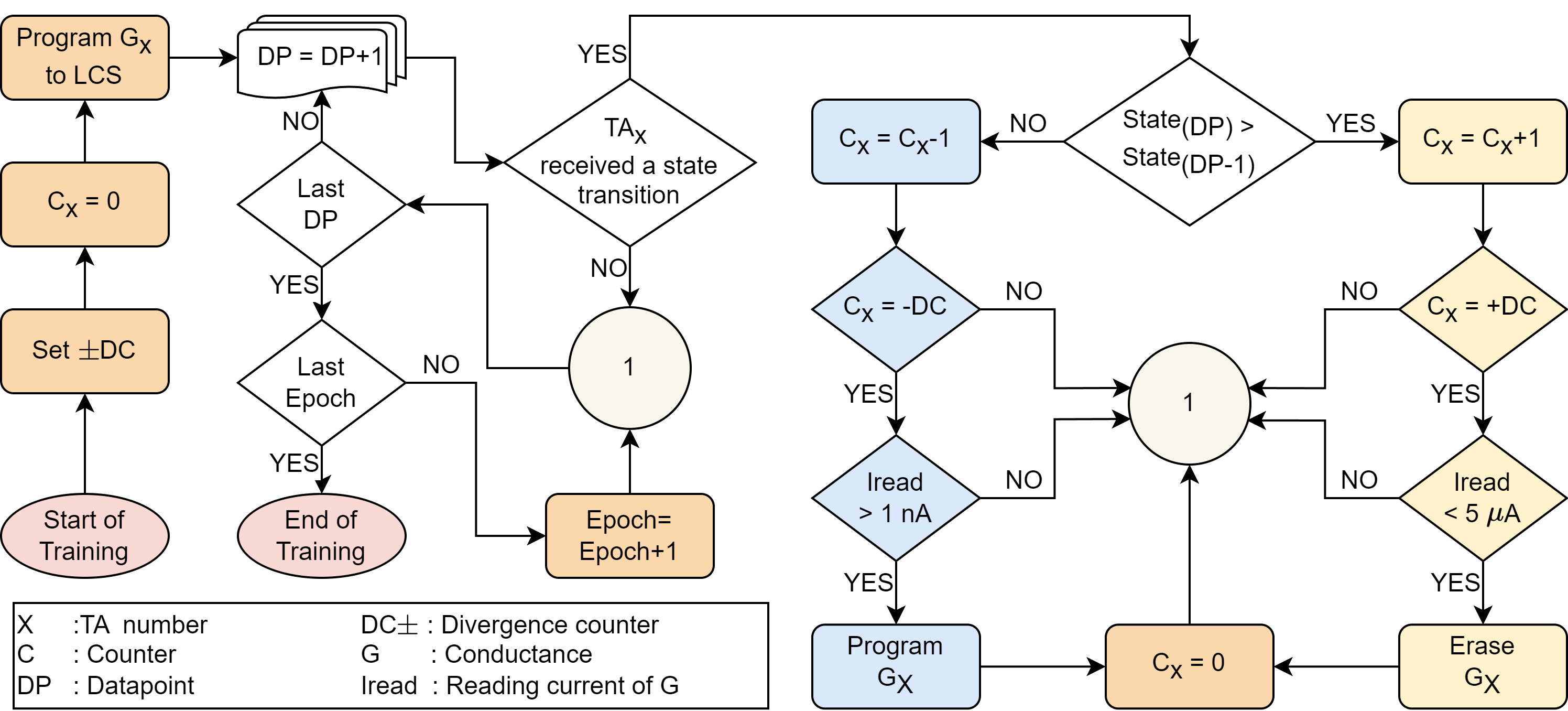}
\caption{ The framework of mimic the TA state transition.}
\label{fig6}
\end{figure}

In Fig.~\ref{fig2}(a, and b), the read currents $(I_{SR})$ of the device are displayed in linear scales. A read voltage $(V_R = 2V)$ was applied with a pulse duration of 5ns. The reading was measured after the device was erased to the high conductance state $(HCS \approx 2.5\mu S)$, which resulted in a reading current of $(I_{SR} \approx 5\mu A)$. The device was then programmed to the low conductance state $(LCS \approx 1nS)$, producing a reading current of $(I_{SR} \approx 1nA)$.The pulse measurements may be affected by overshooting caused by parasitic capacitors in the device. Fig.~\ref{fig3} shows the control of Y-Flash device states, where precise voltage pulses should applied. A program pulse 5V with a 200$\mu s $ width is followed by a $(0-2)V$ sweep to read the device state. Successive programming pulses shift the device from an HCS $(\approx 5\mu A)$  readout current to LCS with a $(\approx 1nA)$. After 40 pulses, the device achieves 41 discrete conductance states. Fig.~\ref{fig3}(a, and b) displays the readout I-V curves before and after each programming operation and the conductance measured at $V_R = 2V$ after each programming pulse, respectively. Similarly, the erase operation is performed by an erase pulse with an amplitude of 8V and a width of 200 $\mu s$. Fig. ~\ref{fig3}(c, and d) displays the readout I-V curves before and after each consecutive erase pulse and conductance measured at $V_R = 2V$ after each erasing pulse. The longer pulses used here mainly speed up the measurement of a complete program cycle. Reducing the program/erase pulse width, like using 10 $\mu s$, can generate over 1000 analog conductance states, providing finer control.
\begin{figure}[tbp]
\centering
\setlength\belowcaptionskip{-1.5\baselineskip}
\includegraphics[width=0.5\textwidth]{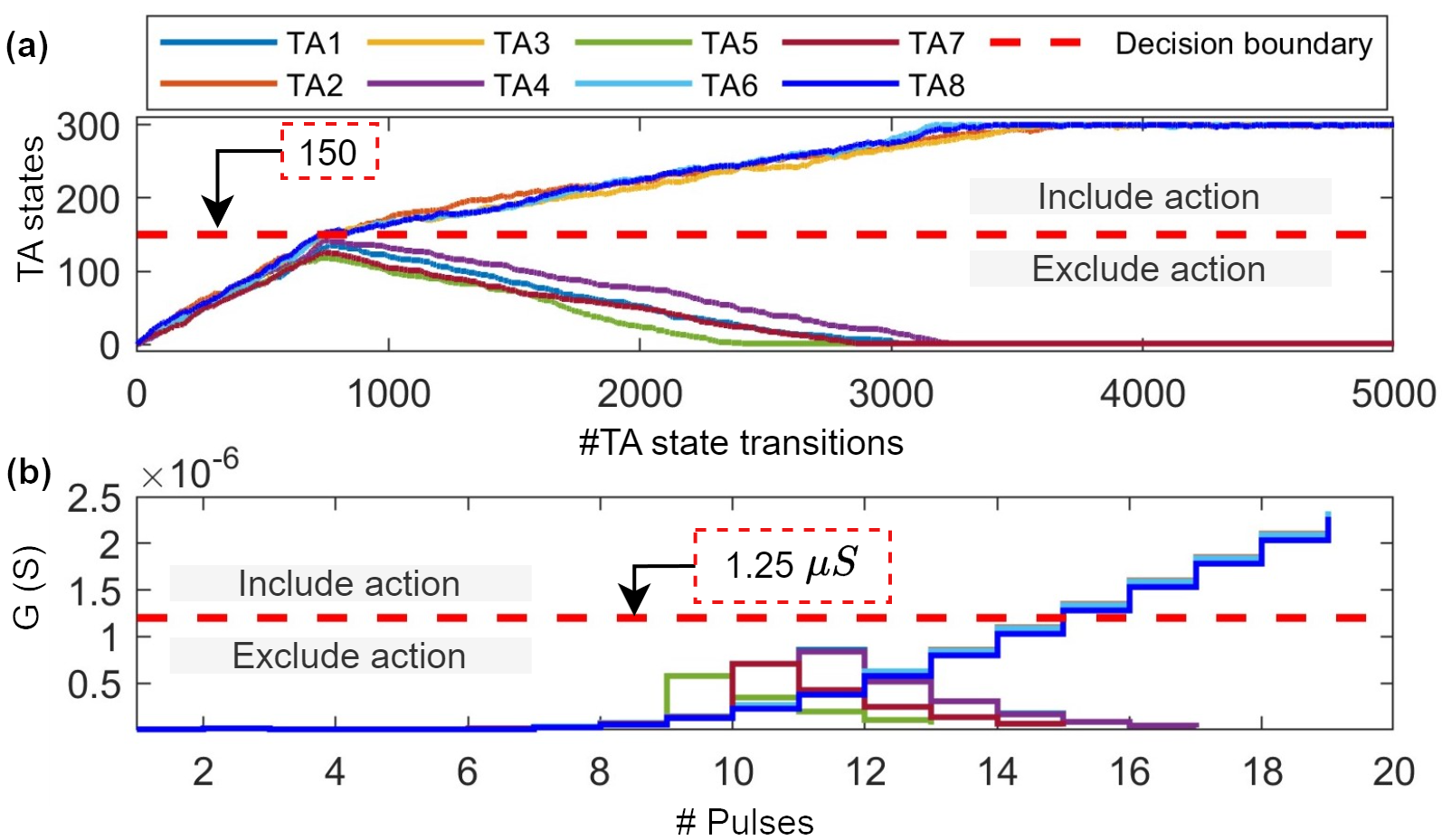}
\caption{(a) State transitions of eight TAs during the learning procedure. (b) Writing cycles of the eight corresponding Y-Flash cells associated with the TAs.}
\label{fig5}
\end{figure}
\subsection{Mapping TM Learning Element to Y-Flash Cell}
This section outlines the methodology for mapping the states of the TA onto individual Y-Flash memory cell. Each distinct state of the TA is translated into a unique conductance state (G) level within the Y-Flash cell, leveraging its ability to operate in multiple analog states. The training procedure, illustrated in Fig.~\ref{fig6}, involves an efficient approach to adjusting the conductance. Instead of directly modifying the conductance based on instantaneous changes in TA states, an accumulating divergence counter \(( \pm\ DC)\) is employed. This counter accumulates the TA state differences over multiple training data points, which mitigates the need for frequent writing to the corresponding Y-Flash cell. When the accumulated DC value surpasses a specified positive or negative threshold (\(\pm\ 15\) in this experiment), a programming or erasing pulse is issued to the corresponding TA to update its conductance. Subsequently, the DC counter is reset to zero, ensuring that the Y-Flash device remains synchronized with the TA's learning dynamics.

To demonstrate the methodology's effectiveness, the TM is trained using the XOR problem dataset. Fig.~\ref{fig5}(a) displays only eight TAs of the TM training set. The decision boundary for the TA was set at 150 states ($2N = 300$). During the training process, which involved 5000 data points, we recorded the state transitions to capture the complexity of the learning dynamics. Notably, four of the TAs reached the 'include' action. Fig.~\ref{fig5}(b) illustrates how our method significantly reduces these transitions to 19 pulses to mimic the TA dynamic transition in the training process, using a pulse width of 0.5 ms, thereby smoothing the mapping of TA state transition behavior. The maximum included TA reached a 2.33 \(\mu S\) conductance, while the minimum excluded TA reached 23.2 nS. This reduction in complexity is achieved while maintaining consistency with the state transitions of the TA. However, using a higher pulse width can realize the TA dynamic transition but with less granularity, necessitating an increase in the limits of the DC. Balancing pulse width and the DC boundaries will determine the required number of pulses, which is crucial for optimizing energy efficiency and programming accuracy. Our approach highlights the efficiency and effectiveness of the method. The integration of the Y-Flash memory cell to represent the multi-conductance states of the TA mimics complex decision-making behavior within a single device. The analog tunable nature of Y-Flash allows for precise adjustment of conductance levels, effectively capturing the accurate learning dynamics of the TA. Another advantage of this approach is utilizing a blind write method, eliminating the need to verify or fine-tune to update the conductance, thereby ensuring a fast and cost-effective write operation. The mapping procedure demonstrates that the seamless integration of TAs within Y-Flash devices holds significant potential for developing efficient and adaptive computing systems capable of handling diverse real-world tasks.

\begin{figure}[tbp]
\centering
\setlength\belowcaptionskip{-1.5\baselineskip}
\includegraphics[width=0.5\textwidth]{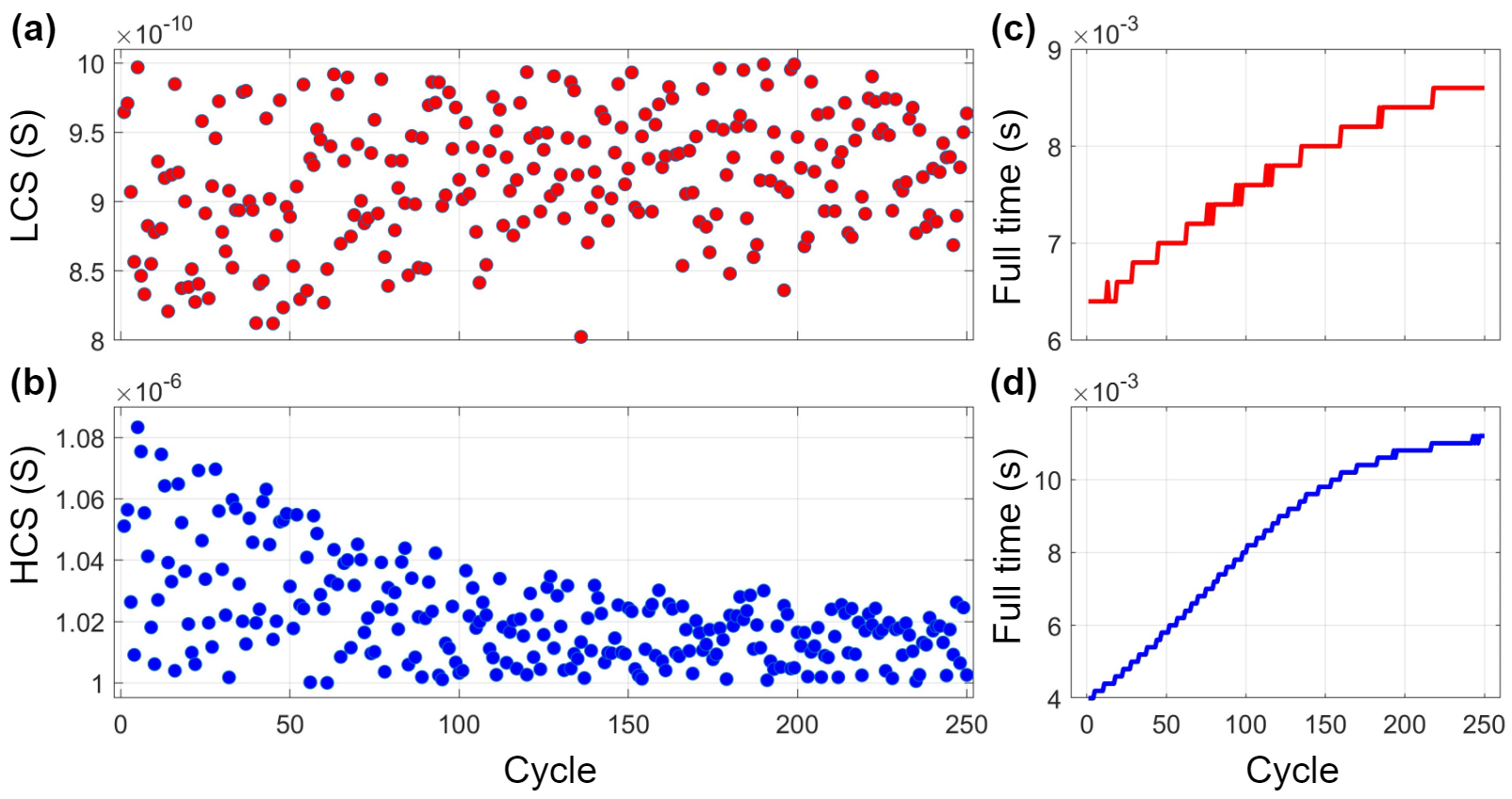}
\caption{(a) and (b) LCS and HCS distributions over 100 cycles, exploring C2C variances. (c) and (d) The full programming and erasing time, respectively.}
\label{fig4}
\end{figure}

\begin{table}[b]
\vspace{-5pt}
\caption{Average consumed power for Y-Flash device.}
\begin{center}
\begin{tabular}{|c|c|c|c|c|}
\hline
\thead{Operation \\ mode} & \thead{Voltage \\ (V)} & \thead{States \\ \#} & \thead{Average \\ power $\mu$W} & \thead{Average \\ energy (nJ)} \\

\hline
Read & 2 & 40 & 1.83  & $9.14 \times 10^{-6}$  \\
\hline
%Program & 5 & 40 & 11.42  & $2.28 \,$ \\
Program & 5 & 40 & \(695\)  & \(139\) \\
\hline
%Erase & 8 & 32 & 59.13  & $1.18 \times 10^{+1} \,$  \\
Erase & 8 & 32 & \(8 \times 10^{-3} \)  & \(1.6 \times 10^{-3} \)  \\
\hline
\end{tabular}
\label{power}
\end{center}

\end{table}

\section{Experiment and Results}
In order to evaluate the conductance of Y-Flash memristors, we performed the cycling performance test C2C. The cycles are illustrated in Fig.~\ref{fig4}(a, and b), using a \(200 \mu s\) pulse width for erase and program operations for 250 cycles. By applying a set of program pulses, we programmed the device at $V_P = 5V$ from HCS \((I_{SR} > 2 \mu A )\) measured at  \( V_R = 2V\) to LCS \((I_{SR} < 1nA)\) and then erased it backward $V_E = 8V$. The results indicated that the devices exhibited a range of LCS between $(0.8\ - \ 0.9)nS$ and HCS between $(1 \ -\  1.08)\mu S$. Fig.~\ref{fig4}(c, and d) show the cycling degradation test through the 250 cycles. The full-time is determined by multiplying the number of pulses by the pulse width. The program time increases in a step-wise manner, reaching a maximum value of \(8.6 \ ms\). Similarly, the erase time increases progressively more noticeably, reaching a maximum value of \(11.2\ ms\). The gradual increase in both program and erase times suggests a slight degradation in device performance over repeated cycling. Despite these increases, no significant performance degradation was observed, as the devices switched between the HCS and LCS reliably over all 250 cycles. The Y-Flash memristors demonstrate robust performance with minimal degradation, maintaining reliable conductance switching throughout the testing period.
 \begin{figure}[tbp]
\centering
\setlength\belowcaptionskip{-1.5\baselineskip}
\includegraphics[width=0.5\textwidth]{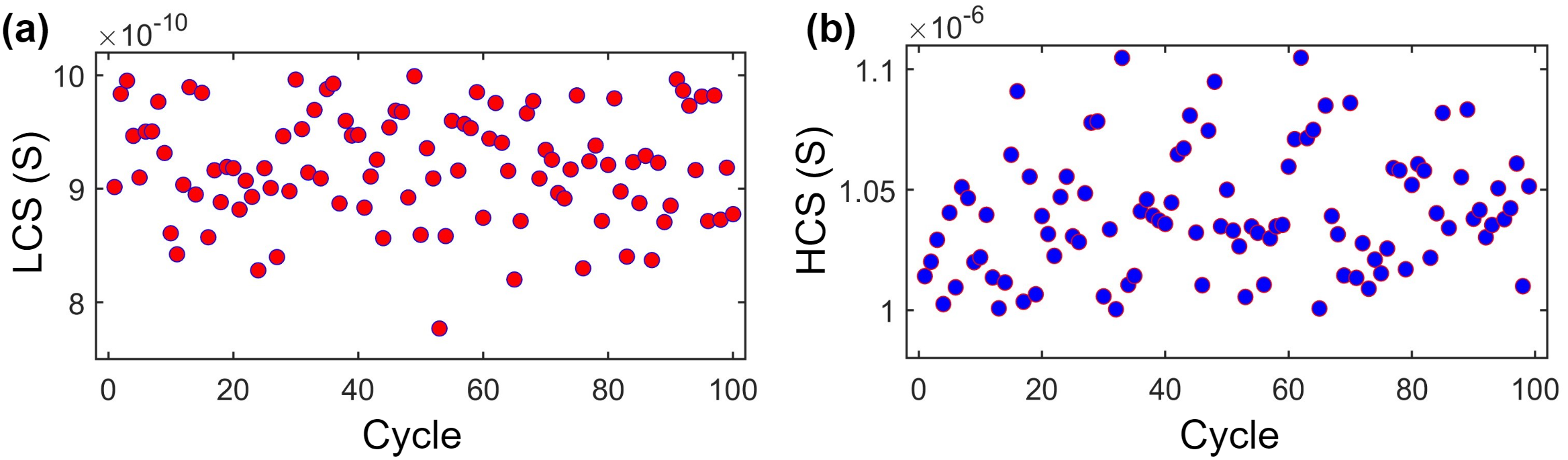}
\caption{D2D variance across 100 devices: (a) Recorded conductances at 2V after programming the devices to LCS. (b) Conductances after erasing the devices to HCS.}
\label{fig7}
\end{figure}
Additionally, we used 100 devices to test D2D variations, as illustrated in Fig.~\ref{fig7}. Following the transition from a target LCS of less than 1 nS to an HCS exceeding 1 \(\mu\)S, as well as the reverse transition, measurements were taken for both states. The LCS exhibited a range between \((0.77 \ - \ 0.99)nS\), while the HCS ranged between \((1.0 \ - \ 1.13) \mu S\). The mean conductance for the LCS was \(0.92 \ nS\) with a standard deviation of \(0.047 \ nS\), whereas for the HCS, the mean conductance was \(1.04 \ \mu S\) with a standard deviation of \(0.027 \ \mu S\). These measurements, plotted against the device number, showed some variance, but all devices functioned normally, indicating a high yield for the tested devices. This success can be attributed to the CMOS fabrication flow and the high uniformity of the analogue switching involved. In Table.\ref{power}, the power calculation was derived based on the 40 conductance states presented in Fig.~\ref{fig3} for both programming and erasing modes, utilizing a pulse duration of 200 $\mu$s to calculate the average energy per mode pulse. Similarly, a pulse duration of 5ns was employed for the reading process, as shown in Fig.~\ref{fig2}.

\section{Conclusion}
This paper presented a new in-memory computing approach utilizing Y-Flash memristive devices, which address data movement and computational efficiency in machine learning architectures while mitigating the von Neumann bottleneck. By representing the learning automaton of the Tsetlin Machine within a single Y-Flash cell, we demonstrate a scalable and energy-efficient hardware implementation that enhances on-edge learning capabilities. Our findings highlight the potential of the Y-Flash memristor to advance the development of reliable in-memory computing for machine learning implementation. Future research will address reliability challenges in memristor devices and optimize the integration of Tsetlin Machines with Y-Flash technology for broader machine-learning applications.

%\section*{Acknowledgment}

\bibliographystyle{IEEEtran}
\bibliography{ref}

\end{document}